\documentclass[sigconf]{acmart}

\AtBeginDocument{%
  \providecommand\BibTeX{{%
    \normalfont B\kern-0.5em{\scshape i\kern-0.25em b}\kern-0.8em\TeX}}}

\copyrightyear{2020}
\acmYear{2020}
\setcopyright{rightsretained}
\acmConference[KDD '20]{Proceedings of the 26th ACM SIGKDD Conference on Knowledge Discovery and Data Mining}{August 23--27, 2020}{Virtual Event, CA, USA}
\acmBooktitle{Proceedings of the 26th ACM SIGKDD Conference on Knowledge Discovery and Data Mining (KDD '20), August 23--27, 2020, Virtual Event, CA, USA}
\acmDOI{10.1145/3394486.3403305}
\acmISBN{978-1-4503-7998-4/20/08}

\usepackage{relsize}
\usepackage{booktabs}
\usepackage{varioref}
\usepackage{adjustbox}
\usepackage{graphicx}
\usepackage{balance}
\begin{document}
\fancyhead{}
\title{Embedding-based Retrieval in Facebook Search}

\author{Jui-Ting Huang}
\email{juiting@fb.com}
\affiliation{
 \institution{Facebook Inc.}
}
\author{Ashish Sharma}
\email{ashishsharma@fb.com}
\affiliation{
 \institution{Facebook Inc.}
}
\author{Shuying Sun}
\email{shuyingsun@fb.com}
\affiliation{
 \institution{Facebook Inc.}
}
\author{Li Xia}
\email{xiali824@fb.com}
\affiliation{
 \institution{Facebook Inc.}
}
\author{David Zhang}
\email{shihaoz@fb.com}
\affiliation{
 \institution{Facebook Inc.}
}
\author{Philip Pronin}
\email{philipp@fb.com}
\affiliation{
 \institution{Facebook Inc.}
}
\author{Janani Padmanabhan}
\email{jananip@fb.com}
\affiliation{
 \institution{Facebook Inc.}
}
\author{Giuseppe Ottaviano}
\email{ott@fb.com}
\affiliation{
 \institution{Facebook Inc.}
}
\author{Linjun Yang}
\authornote{This work was performed when the author was at Facebook.}
\email{yang.linjun@microsoft.com}
\affiliation{
 \institution{Microsoft}
}

\renewcommand{\shortauthors}{Huang et al.}

\begin{abstract}
Search in social networks such as Facebook poses different challenges than in classical web search: besides the query text, it is important to take into account the searcher's context to provide relevant results. Their social graph is an integral part of this context and is a unique aspect of Facebook search.
While embedding-based retrieval (EBR) has been applied in web search engines for years, Facebook search was still mainly based on a Boolean matching model. In this paper, we discuss the techniques for applying EBR to a Facebook Search system. We introduce the unified embedding framework developed to model semantic embeddings for personalized search, and the system to serve embedding-based retrieval in a typical search system based on an inverted index.
We discuss various tricks and experiences on end-to-end optimization of the whole system, including ANN parameter tuning and full-stack optimization. Finally, we present our progress on two selected advanced topics about modeling. We evaluated EBR on verticals\footnote{In Facebook search, verticals are based on result types, e.g., people, page, group, etc.} for Facebook Search with significant metrics gains observed in online A/B experiments. We believe this paper will provide useful insights and experiences to help people on developing embedding-based retrieval systems in search engines.
\end{abstract}



\maketitle

\section{Introduction}
Search engines have been an important tool to help people access the huge amount of information online. Various techniques have been developed to improve search quality in the last decades, especially in web search engines including Bing and Google. Since it is difficult to accurately compute the search intent from query text and represent the semantic meaning of documents, search techniques are mostly based on various term matching methods \cite{mir}, which performs well for the cases that keyword match can address. It still remains a challenging problem for \emph{semantic matching} \cite{Li:2014:SMS}, which is to address desired results that are not exact match of the query text but can satisfy users' search intent.

In the last years, deep learning has made significant progress in speech recognition, computer vision, and natural language understanding \cite{DBLP:journals/nature/LeCunBH15}. Among them \emph{embedding}, which is also called \emph{representation learning}, has been proven to be successful techniques contributing to the success \cite{bengio:representation}. In essence, \emph{embedding} is a way to represent a sparse vector of ids as a dense feature vector, which is also called \emph{semantic embedding} in that it can often learn the semantics. Once the embeddings are learned, it can be used as a representation of query and documents to apply in various stages of a search engine. Due to the huge success of this technique in other domains including computer vision and recommendation system, it has been an active research topic in information retrieval community and search engine industry as the next generation search technology \cite{mitra2018:NeuralIR}.

In general, a search engine comprises a recall layer targeting to retrieve a set of relevant documents in low latency and computational cost, usually called \emph{retrieval} , and a precision layer targeting to rank the most desired documents on the top with more complex algorithms or models, usually called \emph{ranking}. While embeddings can be applied to both layers, it usually has more opportunities to leverage embeddings in the retrieval layer, since it is at the bottom of the system which is often the bottleneck. The application of embeddings in retrieval is called \emph{embedding-based retrieval} or \emph{EBR} for short. Briefly, embedding-based retrieval is a technique to use embeddings to represent query and documents, and then convert the retrieval problem into a nearest neighbor (NN) search problem in the embedding space.

EBR is a challenging problem in search engines because of the huge scale of data being considered. Different from ranking layers which usually takes hundreds of documents into consideration per session, retrieval layer needs to process billions or trillions of documents in the index of a search engine. The huge scale imposes challenges on both training of embeddings and serving of embeddings. Second, different from embedding-based retrieval in computer vision tasks, search engine usually needs to incorporate both embedding-based retrieval and term matching based retrieval together to score documents in the retrieval layer.

Facebook search, as a social search engine, has unique challenges compared with traditional search engines. In Facebook search, the search intent does not only depend on query text but is also heavily influenced by the user who is issuing the query and the context where the searcher is. Because of this, embedding-based retrieval in Facebook search is not a text embedding problem, as is actively researched in the IR community \cite{mitra2018:NeuralIR}. Instead it is a more complex problem that requires understanding of text, user, and the context altogether.

To deploy embedding-based retrieval in Facebook search, we developed approaches to address challenges on \emph{modeling}, \emph{serving}, and \emph{full-stack optimization}. In modeling, we proposed \emph{unified embedding}, which is a two sided model where one side is search request comprising query text, searcher, and context, and the other side is the document. To effectively train the model, we developed approaches to mine training data from search log and extract features from searcher, query, context, and documents. For fast model iteration, we adopted a recall metric on an offline evaluation set to compare models.

Building retrieval models for search engine has its unique challenges, such as how to build a representative training task for models to learn effectively and efficiently. We investigated two different directions, \emph{hard mining} to address the challenge of representing and learning retrieval tasks effectively, as well as \emph{ensemble embedding} to divide the model in multiple stages where each stage has different recall and precision tradeoff.

After the model is developed, we need to develop ways to effectively and efficiently serve the model in the retrieval stack. While it is straightforward to build a system combining the candidates from existing retrieval and embedding KNN, we found it is suboptimal because of several reasons: 1) it has huge performance cost from our initial experiment; 2) there is high maintenance cost because of dual index; 3) the two candidate sets might have significant overlap which makes it inefficient overall. Thereafter, we developed a hybrid retrieval framework to integrate embedding KNN and Boolean matching together to score documents for retrieval. To this purpose, we employed Faiss \cite{JDH17} library for embedding vector quantization and integrated it with inverted index based retrieval to build a hybrid retrieval system. Besides addressing the above challenges, this system has two main advantages: 1) it enables the joint optimization of embedding and term matching to address Search retrieval problem; 2) it supports embedding KNN constrained by term matching, which not only helps address the system performance cost issue but also improves the precision of embedding KNN results.

\begin{figure}
\includegraphics[scale=0.4]{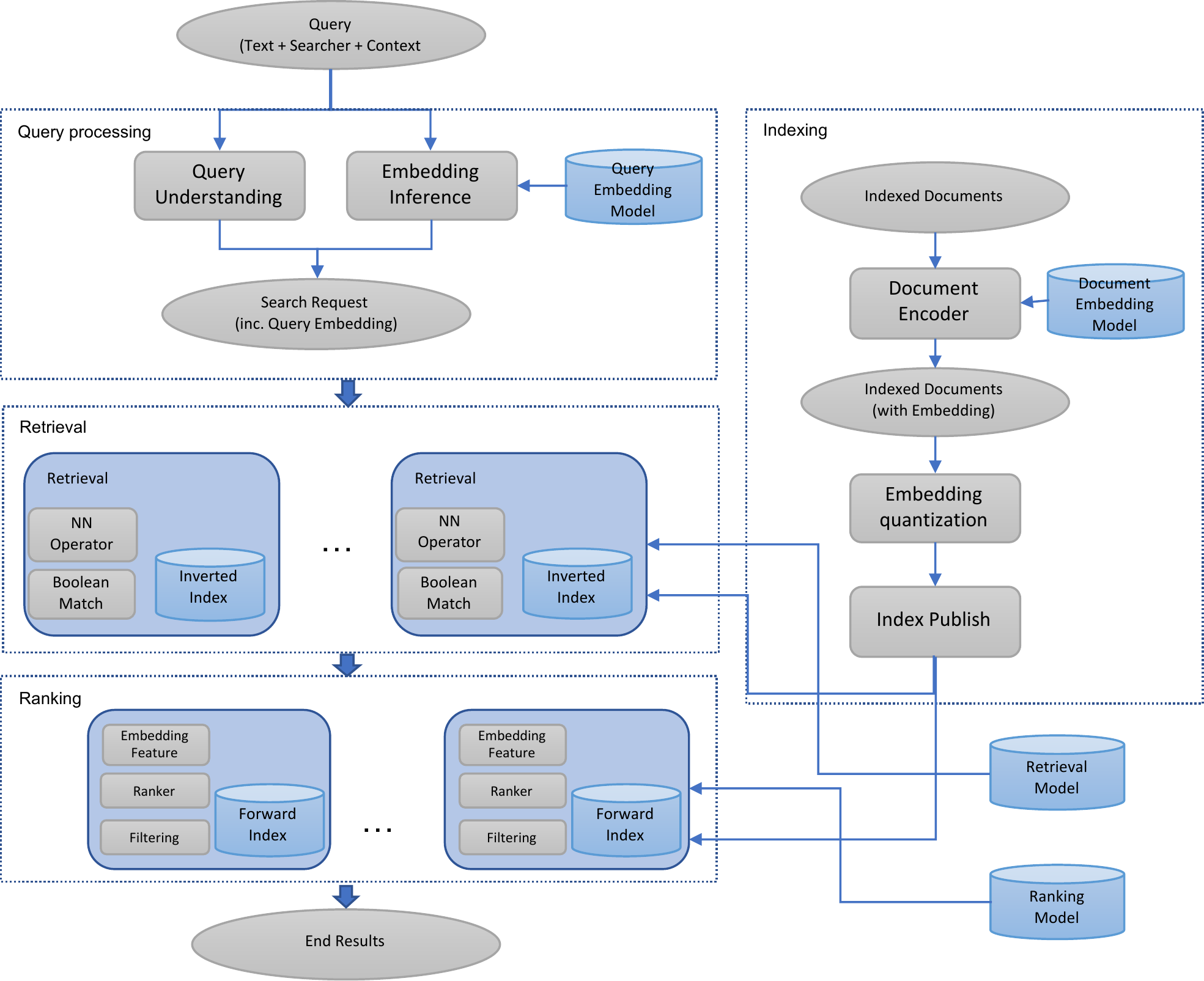}
\caption{Embedding Based Retrieval System Overview}
\label{fig:EBR_System_Overview}
\end{figure}

Search is a multi-stage ranking system where retrieval is the first stage, followed by various stages of ranking and filtering models. To wholly optimize the system to return those new good results and suppress those new bad results in the end, we performed later-stage optimization. In particular, we incorporated embeddings into ranking layers and built a training data feedback loop to actively learn to identify those good and bad results from embedding-based retrieval. Figure~\ref{fig:EBR_System_Overview} is an illustration of embedding-based retrieval system. We evaluated EBR on verticals for Facebook Search with significant metrics gains observed in online A/B experiments.

The paper is organized as follows. We start with modeling to present our solutions about loss function, model architecture, training data and feature engineering in Section~\ref{sec:model} and Section~\ref{sec:feature}. Next we discuss about details on model serving and system implementation in Section \ref{sec:serving}. We discuss the techniques we developed on later-stage optimization to unleash the power from embedding-based retrieval end to end in Section~\ref{sec:fullstack}. Finally, we dedicate Section~\ref{sec:advanced} to selected topics on advanced modeling techniques, followed by conclusions in Section \ref{sec:conclusion}.

\section{Model}
\label{sec:model}
We formulate the search retrieval task as a recall optimization problem. Specifically, given a search query, its target result set $T=\{t_1,t_2,...t_N\}$, and top $K$ results returned by a model, $\{d_1,d_2,...d_K\}$, we want to maximize recall by the top $K$ results,
\begin{equation}\label{eq:recall}
  recall@K=\frac{\sum_{i=1}^{K}d_i\in T}{N}.
\end{equation}
The target results are the documents related to the given query based on certain criteria. For example, it could be results with user clicks, or relevant documents based on human rating.

We formulate the recall optimization as a ranking problem based on distances computed between a query and documents. The query and documents are encoded with a neural network model into dense vectors, on which we use cosine similarity as the distance metric. We propose to use triplet loss \cite{FaceNet} to approximate the recall objective to learn the neural network encoder, which is also called embedding model.

While semantic embedding is commonly formulated as text embedding problem in information retrieval, it is insufficient for Facebook search, which is a personalized search engine that considers not only text query but also searcher's information as well as the context in a search task to satisfy users' personalized information need. Taking people search as an example, while there might be thousands user profiles named "John Smith" on Facebook, the actual target person that a user searches for with query "John Smith" is likely to be their friends or acquaintances. To model this problem, we propose \emph{unified embedding} which considers not only text but also user and context information in deriving embeddings.

\subsection{Evaluation Metrics}
\label{sec:eval}
While our end goal is to deliver quality improvement end to end through online A/B test, it is important to develop offline metrics to quickly evaluate model quality before online experiments and isolate problems from complicated online experiment setup. We propose to run KNN search in the whole index and then use \emph{recall@K} as defined in equation~\ref{eq:recall} as the model evaluation metric. In particular, we sampled 10000 search sessions to gather the query and target result set pairs for the evaluation set and reported averaged \emph{recall@K} over 10000 sessions.

\subsection{Loss Function}
\label{sec:loss}
For a given triplet $(q^{(i)},d^{(i)}_{+},d^{(i)}_{-})$, where $q^{(i)}$ is a query, $d^{(i)}_{+}$ and $d^{(i)}_{-}$ are the associated positive and negative documents, respectively, the triplet loss is defined as
\begin{equation}\label{eq:loss}
L= \sum_{i=1}^{N}\max(0, D(q^{(i)},d^{(i)}_{+}) - D(q^{(i)},d^{(i)}_{-}) + m),
\end{equation}
where $D(u,v)$ is a distance metric between vector $u$ and $v$, $m$ is the margin enforced between positive and negative pairs, and $N$ is the total number of triplets selected from the training set. The intuition of this loss function is to separate the positive pair from the negative pair by a distance margin. We found that tuning margin value is important -- the optimal margin value varies a lot across different training tasks, and different margin values result in 5-10\% KNN recall variance.

We believe that using random samples to form negative pairs for the triplet loss can approximate the recall optimization task. The reason is as follows. If we sample $n$ negatives for each one positive in the training data, the model will be optimizing for recall at \emph{top one position} when the candidate pool size is $n$. Assuming the actual serving candidate pool size is $N$, we are approximately optimizing recall at top $K \approx N/n$. In Section~\ref{sec:data}, we will verify this hypothesis and provide comparisons of different positive and negative label definitions.

\subsection{Unified Embedding Model}
To learn embeddings that are optimizing the triplet loss, our model comprises three major components: a query encoder $E_Q=f(Q)$ which produces a query embedding, a document encoder $E_D=g(D)$ which produces a document embedding, and a similarity function $S(E_Q,E_D)$ which produces a score between query $Q$ and document $D$.
An encoder is a neural network which transforms an input into a low-dimensional dense vector, also known as embedding. In our model, these two encoders $f(\cdot)$ and $g(\cdot)$ by default are two separate networks but have the option of sharing part of the parameters. As for similarity function, we choose cosine similarity as it is one of the commonly used in embedding learning \cite{DSSM}:
\begin{equation}
S(Q,D)=\cos(E_Q,E_D)=\frac{\langle E_Q,\ E_D\rangle}{\|E_Q\|\cdot\|E_D\|}.
\end{equation}
The distance to be used in the loss function in Equation~\ref{eq:loss} is hence the cosine distance defined as $1-\cos(E_Q,E_D)$.

The inputs to the encoders are what distinguishes unified embedding from conventional text embedding model. Unified embedding encodes textual, social and other meaningful contextual features to represent query and document, respectively. For example, for query side we can include searcher location and their social connections, whereas for the document side we can include aggregated location and social clusters about a Facebook group by taking groups search as an example.

\begin{figure}

  \includegraphics[width=0.45\textwidth]{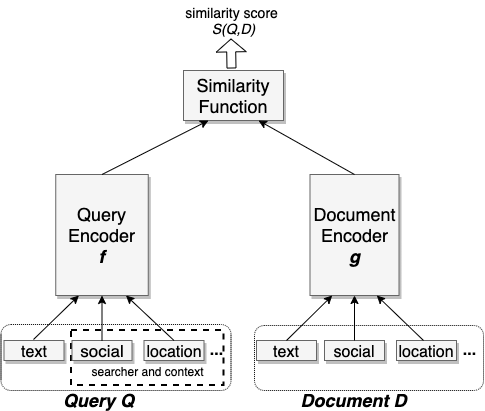}

  \caption{Unified Embedding Model Architecture}

  \label{fig:ttsn}

\end{figure}

Most of features are categorical features of high cardinality, which could be either one-hot or multi-hot vectors. For each categorical feature, an embedding look-up layer is inserted to learn and output its dense vector representation before feeding into encoders. For multi-hot vectors, a weighted combination of multiple embeddings is applied for the final feature-level embedding. Figure~\ref{fig:ttsn} illustrates our unified embedding model architecture, and we will discuss more about feature engineering in Section~\ref{sec:feature}.

\subsection{Training Data Mining}

\label{sec:data}

Defining positive and negative labels for a retrieval task in a search ranking system is a non-trivial problem. Here we compared several options based on the model recall metric. For negative, we experimented with the following two options of negatives in our initial study, while using click as positive:
\begin{itemize}
  \item \emph{random samples}: for each query, we randomly sample documents from the document pool as negatives.
  \item \emph{non-click impressions}: for each query, we randomly sample those impressed but not clicked results in the same session as negatives.
\end{itemize}
The model trained using non-click impressions as negative has significantly worse model recall compared to using random negative: absolute 55\% regression in recall for people embedding model.
We believe it is because these negatives bias towards hard cases which might match the query in one or multiple factors, while the majority of documents in index are easy cases which do not match the query at all. Having all negatives being such hard negatives will change the representativeness of the training data to the real retrieval task, which might impose non-trivial bias to the learned embeddings.

We also experimented with different ways of mining positives and have interesting findings as follows:
\begin{itemize}
  \item \emph{clicks}: it is intuitive to use clicked results as positives, since clicks indicates users' feedback of the result being a likely match to users' search intent.
  \item \emph{impressions}: the idea is that we treat retrieval as an approximation to ranker but can execute fast. Thereafter, we want to design the retrieval model to learn to return the same set of results that will be ranked high by the ranker. In this sense, all results shown or impressed to the users are equally positive for retrieval model learning.
\end{itemize}
Our experimental results showed that both definitions are equally effective; models trained using click vs impressions, given the same data volume, resulted in similar recalls. Additionally, we experimented with augmenting click-based training data with impression-based data, however we did not observe additional gain over the click-based model. It showed that adding impression data does not provide additional value, and the model does not benefit from increased training data volume either.

Our above study suggested that using click as positive and random as negative can provide a reasonable model performance. On top of it, we further explored hard mining strategies to improve the model's ability of differentiating between similar results. We will present more details in Section~\ref{hard}.

\section{Feature Engineering}
\label{sec:feature}
One of the advantages of unified embedding model is that it can incorporate various features other than text to improve the model performance. We observed consistently across different verticals that unified embedding is more effective than text embedding.
For example, there are +18\% recall improvement when switching from text to unified embeddings for events search, and +16\% recall improvement for groups search. The effectiveness of unified embeddings highly depends on the success of identifying and crafting informative features. Table~\ref{tab:recall} shows the incremental improvement by adding each new feature category to the group embedding model (with text features as baseline). In this section we discuss several important features that contributed to the major model improvements.

\begin{table}
  \caption{Group Embedding Improvement with Feature Engineering}
  \centering
  \begin{adjustbox}{center}
  \begin{tabular}{@{}l|l@{}}
  \toprule
  \textbf{Unified Embedding}  & \textbf{Abs. Recall Gain} \\ \midrule
  + location features         & + 2.20\%                     \\
    \ \ \ + social embedding features & + 1.77\%                     \\ \bottomrule
  \end{tabular}
  \end{adjustbox}

  \label{tab:recall}
\end{table}

\textbf{Text features.} Character n-gram \cite{DSSM} is a common approach to represent text for text embedding. Its advantages compared to word n-grams are two folds. First, due to its limited vocabulary size, the embedding lookup table has a smaller size and can be learned more effectively during training. Second, the subword representation is robust to out-of-vocabulary problem the we encounter for both query (e.g. spelling variations or errors) and document sides (due to large content inventory in Facebook). We compared models trained with character n-grams vs word n-grams and found the former can yield a better model. However, on top of character trigrams, including word n-gram representations additionally provides small but consistent model improvement (+1.5\% recall gain). Note that since the cardinality of word n-grams is usually very high (e.g. 352M for query trigrams), hashing is needed to reduce the size of embedding lookup table. But even with the downside of hash collision, adding word n-ngrams still provides extra gain.

For a Facebook entity\footnote{Facebook entity includes people (profile), group, page, and event.}, the main field to extract text features is name for people entities or title for non-people entities. Compared to Boolean term matching techniques, we found embeddings trained with purely text features are particularly good at addressing the following two scenarios:

\begin{itemize}
  \item Fuzzy text match. For example, the model is able to learn to match between query "kacis creations" and the \textit{Kasie's creations} page while the term-based match cannot.
  \item Optionalization. In the case of query "mini cooper nw", the model can learn to retrieve the expected group \textit{Mini cooper owner/drivers club} by dropping "nw" for an optional term match.
\end{itemize}

  \begin{table}

    \centering
    \caption{Top Similar Groups Before and After Adding Location Embeddings}

    \begin{adjustbox}{width=\columnwidth,center}

    \begin{tabular}{@{}llll@{}}

    \toprule

    \multicolumn{4}{c}{\textbf{Searcher Location}: Louisville, Kentucky}\\
    \multicolumn{4}{c}{\textbf{Query}: \textit{"equipment for sale"}}\\
    \midrule
\multicolumn{2}{c}{\textbf{Text Model}} & \multicolumn{2}{c}{\textbf{Text + Location Model}} \\
    \midrule

    1 & equipment for sale &1 &Kentucky Farm Equipment for sale                                            \\ 

    2 & EQUIPMENT FOR SALE WORLDWIDE &2& \begin{tabular}[c]{@{}l@{}}Pre-Owned Farm Equipment for Sale in KY, \\ Southern Indiana and Tennessee\end{tabular} \\ 

    3 & EQUIPMENT SALE &3 &Farm Equipment For Sale In Ky \\ 

    4 & Equipment for Sale or Wanted &4 &Central Ky Farm Equipment For Sale \\ 

    5 & Musical Equipment For Sale or Trade &5 &sed Farm Equipment for sale or trade East Ky. \\ 

    6 & Used Heavy Equipment For Sale in US &6 &Farm Equipment KY for Sale \\ 

    7 & Sale equipment & 7&kentucky hay and farm equipment for sale \\ \bottomrule

    \end{tabular}

    \end{adjustbox}

    \label{tab:s2s}

    \end{table}

\textbf{Location features.}
Location match is advantageous in many search scenarios such as searching for local business/groups/events. In order for embedding models to consider locations in generating the output embeddings, we added location features into both query and document side features. For query side, we extracted searcher city, region, country, and language. For document side, we added publicly available information, such as explicit group location tagged by the admin. Together with text features, the model was able to successfully learn implicit location match between query and results. Table~\ref{tab:s2s} shows a side-by-side comparison of top similar documents returned by text embedding model versus text + location embedding model for groups search. We can see the model with location features can learn to fuse location signals into embeddings, ranking documents that has the same location as the searcher who is from Louisville, Kentucky to higher positions.

\textbf{Social embedding features.}
To leverage the rich Facebook social graph to improve unified embedding model, we trained a separate embedding model to embed users and entities based on the social graph. This can help incorporate the comprehensive social graph into a unified embedding model which may not have full social graph information otherwise.

\section{Serving}
\label{sec:serving}
\subsection{ANN}
We deployed an inverted index based ANN (approximate near neighbor) search algorithms to our system because of the following advantages. First, it has smaller storage cost due to the quantization of embedding vectors. Second, it is easier to be integrated into the existing retrieval system which is based on inverted index. We employed Faiss library \cite{JDH17} to quantize the vectors and then implemented the efficient NN search in our existing inverted table scanning system.

There are two major components for the embedding quantization, one is the \emph{coarse quantization} which quantizes embedding vectors into  coarse clusters typically through \emph{K-means} algorithm, and the other is \emph{product quantization} \cite{PQ} which does a fine-grained quantization to enable efficient calculation of embedding distances. There are several important parameters we need to tune:

\begin{itemize}

  \item \emph{Coarse quantization.} There are different algorithms for coarse quantization. It is useful to compare between IMI \cite{IMI} and IVF \cite{IVF} algorithms. And it is important to tune number of coarse clusters \emph{num\_cluster}, which will affect both perf and recall.

  \item \emph{Product quantization.} There are multiple variants of product quantization algorithms, including vanilla PQ, OPQ, PQ with PCA transform. And number of bytes for PQ \emph{pq\_bytes} is an important parameter to tune.

  \item \emph{nprobe.} \emph{nprobe} is the parameter to decide how many clusters will be assigned to the query embedding, which will further decide how many coarse clusters will be scanned. This parameter will affect the perf and recall.

\end{itemize}
\begin{figure}

  \includegraphics[scale=0.7]{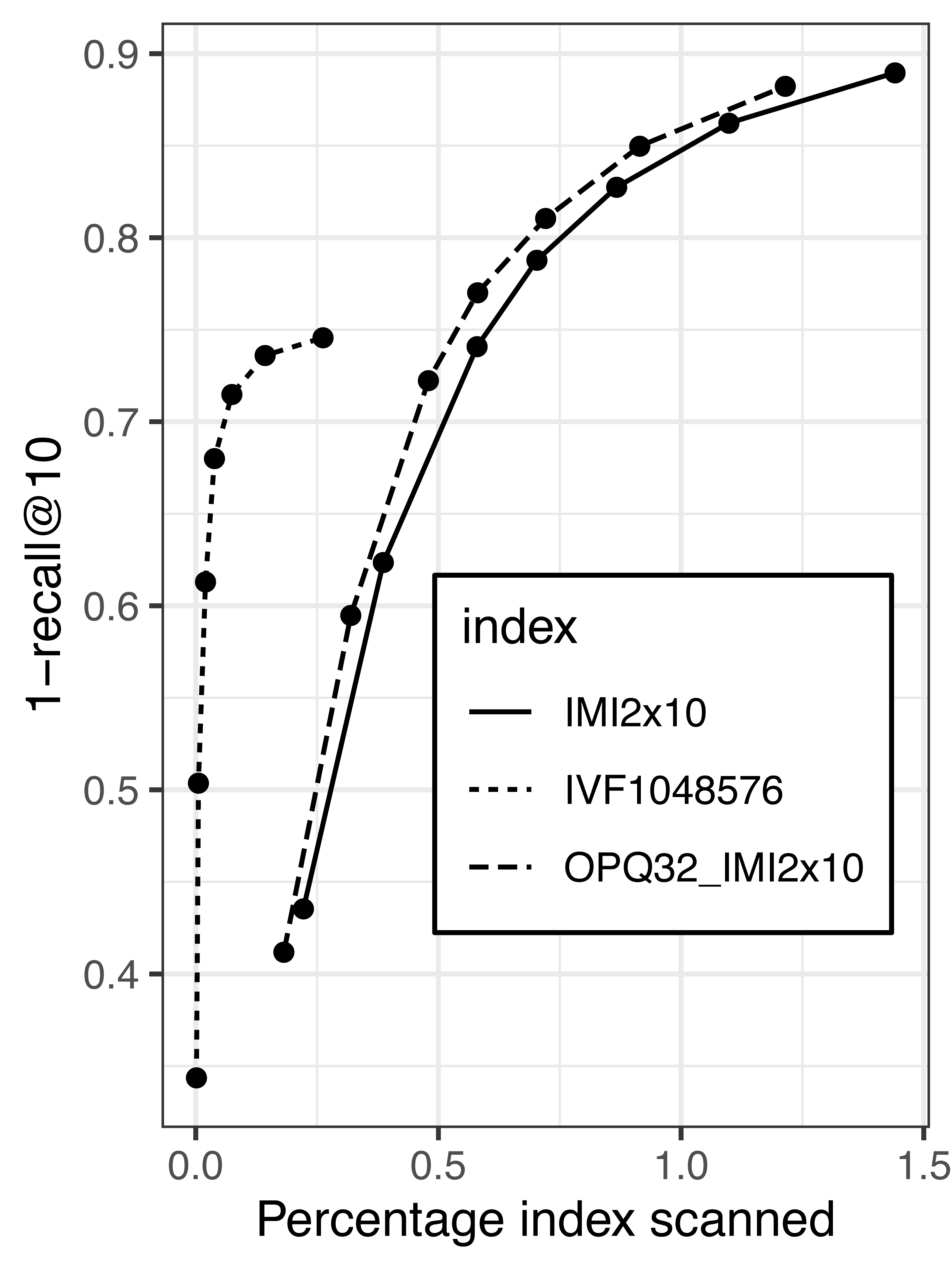}

  \caption{1-Recall@10 of IVF and IMI Index}

  \label{fig:IMIvsIVF-big}

\end{figure}

\begin{table}[h]
  \caption{Impact of Product Quantization on 1-Recall@10 for 128-Dimension Embedding.}
  \centering

  \begin{adjustbox}{width=\columnwidth,center}

  \begin{tabular}{lll}

    \toprule

    \textbf{Quantization Scheme}    & \textbf{1-Recall@10}  & \textbf{Index Scanned}               \\

    \midrule
    PCA,PQ16 & 51.62\% & 0.48\% \\
    PCA,PQ16  & 54.11\% & 0.60\% \\
    PQ16 & 67.54\% & 0.58\%\\
    OPQ16,PQ16 & 70.51\% & 0.48\% \\
    OPQ16,PQ16 & 74.29\% & 0.60\% \\
    PQ32 & 74.27\% & 0.58\%\\
    PQ64 & 74.81\% & 0.58\%\\
    Flat (No Quantization) & 74.81\% & 0.58\%\\

    \bottomrule

  \end{tabular}

  \end{adjustbox}

  \label{tab:pq}

\end{table}

We built an offline pipeline to efficiently tune these parameters. Besides, we needed to run online experiment to decide the final setting from the selected candidates based on offline tuning. Below we will share the tricks and learnings we got from ANN tuning.

\begin{itemize}

\item \emph{Tune recall against number of scanned documents.} Initially we compared recall of different coarse quantization algorithms for the same setting of \emph{num\_cluster} and \emph{nprobe}. However, from more data analysis we found that the clusters are imbalanced, especially for IMI algorithm -- around half of clusters only got a few samples. This would cause the number of scanned documents different for the same setting of \emph{num\_cluster} and \emph{nprobe}. Therefore, we employed the number of scanned documents as a better metric to approximate perf impact in ANN tuning, as shown in Figure \ref{fig:IMIvsIVF-big}. We measured ANN accuracy using \emph{1-recall@10}, which is averaged recall of getting top result from exact KNN search in top 10 results of ANN search.

\item \emph{Tune ANN parameters when there is non-trivial model change.}  We observed that ANN performance is related with model characteristics. For example, when we employed ensemble techniques with model trained with non-click impressions, we found that while the model showed a better recall than baseline, the recall was worse than baseline after applying quantization to both. It should always be considered to tune the ANN parameters when there is a non-trivial change of the model training task, e.g., add more hard negatives.

\item \emph{Always try OPQ.}
It is often useful to transform data prior to applying the quantization. We experimented with both PCA and OPQ \cite{Ge_2013_CVPR} to transform the data, and observed that OPQ is generally more effective, as shown in Table \ref{tab:pq} and Figure \ref{fig:IMIvsIVF-big}. One caveat of OPQ is: as it applied rotation of the embeddings, we might also need to retune \emph{num\_cluster} and \emph{nprobe} to have similar documents scanned.

\item \emph{Choose \emph{pq\_bytes} to be d/4.}
Product quantizer compresses the vectors into $x$ byte codes. For the choice of $x$, it is related to the dimension $d$ of the embedding vector. Bigger $x$ results in higher search accuracy but at cost of increased memory and latency. From empirical results, we found that the accuracy improvement is limited after $x > d/4$.

\item \emph{Tune nprobe, num\_clusters, and pq\_bytes online to understand the real perf impact.} While it is important to tune ANN algorithms and parameters offline to get a reasonable understanding of perf vs. recall trade-off, we found it is important to deploy several configs of the ANN algorithms and parameters online to get a better understanding of the perf impact from embedding-based retrieval to the real system. That is important to decide the capacity budget and reduce the scope of parameter search in offline tuning.

\end{itemize}

\subsection{System Implementation}
In order to integrate embedding-based retrieval into our serving stack, we implemented first-class support for NN search in Unicorn~\cite{Unicorn13}, a retrieval engine powering most search products at Facebook. Unicorn represents each document as a bag of terms, which are arbitrary strings that express binary properties about the document, conventionally namespaced with their semantics. For example, a user \emph{John} living in \emph{Seattle} would have the terms \verb|text:john| and \verb|location:seattle|. Terms can have payloads attached to them.

A query can be any Boolean expression on the terms. For example, the following query would return all the people that have \emph{john} and \emph{smithe} in the name, and live in \emph{Seattle} or \emph{Menlo Park}:

\begin{verbatim}
(and (or (term location:seattle)
         (term location:menlo_park))
     (and (term text:john)
          (term text:smithe)))
\end{verbatim}

To support NN, we extended the document representation to include embeddings, each with a given string key, and added a \verb|(nn <key> :radius <radius>)| query operator which matches all documents whose \verb|<key>| embedding is within the specified radius of the query embedding.

At indexing time, each document embedding is quantized and turned into a term (for its coarse cluster) and a payload (for the quantized residual). At query time, the \verb|(nn)| is internally rewritten into an \verb|(or)| of the terms associated to the coarse clusters closest to the query embedding (\emph{probes}), and for matching documents the term payload is retrieved to verify the radius constraint. The number of probes can be specified with an additional attribute \verb|:nprobe|.
By implementing NN support in terms of pre-existing primitives, instead of writing a separate system, we inherited all the features of the existing system, such as realtime updates, efficient query planning and execution, and support for multi-hop queries (see~\cite{Unicorn13}).

The latter allows us to support top-K NN queries, where instead of matching by radius we select only the K documents closest to the query, and then evaluate the rest of the query. However, from our experimental study, we found that radius mode can give better trade-off of system performance and result quality. One possible reason is that radius mode enables a constrained NN search (constrained by other parts of the matching expression) but top $K$ mode provides a more relaxed operation which needs to scan the whole index to get top $K$ results. Hence, we use radius based matching in our current production.

\subsubsection{Hybrid Retrieval}
By having the \verb|(nn)| operator as part of our Boolean query language we can now support \emph{hybrid} retrieval expressions, with arbitrary combinations of embeddings and terms. This can be used for model based fuzzy matching which can improve on cases like spelling variations, optionalization etc. while reusing and benefiting from other parts of retrieval expression. For example, say a mis-spelled query \emph{john smithe} is looking for a person named \emph{john smith} in \emph{Seattle} or \emph{Menlo Park}; the retrieval expression would look like the one above.

That expression will fail to retrieve the user in question since the term \verb|text:smithe| will fail to match that document. We can add fuzzy matching to this expression through \verb|(nn)| operator:
\begin{verbatim}
(and (or (term location:seattle)
         (term location:menlo_park))
     (or (and (term text:john)
              (term text:smithe))
         (nn model-141795009 :radius 0.24 :nprobe 16)))
\end{verbatim}
where \verb|model-141795009| is the key for the embedding. In this case the target user will be retrieved if the cosine distance between the query (\emph{john smithe}) embedding and document (\emph{john smith}) embedding is less than 0.24.

\subsubsection{Model Serving}
We served the embedding model in the following way. After the two-sided embedding model was trained, we decomposed the model into a query embedding model and a document embedding model and then served the two models separately. For query embedding, we deployed the model in an online embedding inference service for real-time inference. For documents, we used Spark to do model inference in batch offline, and then published the generated embeddings together with other metadata into forward index. We did additional embedding quantization including coarse quantization and PQ to publish it into inverted index.

\subsection{Query and Index Selection}

To improve efficiency and quality of EBR, we performed query and index selection. We applied the query selection technique to overcome problems like over-triggering, huge capacity cost and junkiness increase. We did not trigger EBR for certain queries as EBR would be poor at and provide no extra value for them, such as easy queries with which searchers are looking for a specific target searched and clicked before, or queries with clearly different query intents from what the embedding model was trained for. On index side, we did index selection to make searching faster. For example, we only chose monthly active users, recent events, popular pages and groups.

\section{Later-stage Optimization}
\label{sec:fullstack}
Facebook search ranking is a complex multi-stage ranking system where each stage progressively refines the results from the preceding stage. At the very bottom of this stack is the retrieval layer, where embedding based retrieval is applied. Results from the retrieval layer are then sorted and filtered by a stack of ranking layers. The model at each stage should be optimized for the distribution of results returned by the preceding layer. However, since the current ranking stages are designed for existing retrieval scenarios, this could result in new results returned from embedding based retrieval to be ranked sub-optimally by the existing rankers. To solve this problem, we proposed two approaches:

\begin{itemize}
\item \emph{Embedding} as ranking feature. Propagating embedding similarities further down the funnel not only helps the ranker recognize new results from embedding-based retrieval, but also provides a generic semantic similarity measure for all results. We explored several options to extract features based on embeddings, including cosine similarity between the query and result embeddings, Hadamard product, and raw embeddings. From our experimental studies, cosine similarity feature consistently showed better performance than other options.

\item \emph{Training data feedback loop}. While embedding-based retrieval can improve retrieval recall, it might have a lower precision in comparison with term matching. To address the precision issue, we built a closed feedback loop based on human rating pipeline. In particular, we logged the results after enabling embedding-based retrieval, and then sent these results to human raters to label whether they are relevant or not. We used these human rated data to re-train the relevance model so that it can be used to filter out the irrelevant results from embedding-based retrieval while keeping the relevant ones. This proved to be a useful technique to achieve high precision for the recall improvement in embedding-based retrieval.

\end{itemize}

\section{Advanced Topics}
Embedding-based retrieval requires extensive research to continue improve the performance. We investigated two important areas for embedding modeling: hard mining and embedding ensemble, to continue advance the state of the art of embedding-based retrieval solutions.
\label{sec:advanced}
\subsection{Hard Mining} \label{hard}
The data space for a retrieval task has diverse data distribution in degrees of text/semantic/social matches, and it is important to design a training data set for an embedding model to learn efficiently and effectively on such space.
To tackle this, hard mining is one major direction as well as an active research area for embedding learning. However, most of the research are from computer vision field and for the classification task \cite{FaceNet,DBLP:journals/corr/SongXJS15,DBLP:journals/corr/HermansBL17,DBLP:journals/corr/WuMSK17}, while search retrieval does not have concept of "classes" and therefore is a unique problem for which existing techniques do not necessarily work.
In this direction, we divided our solutions into two parts: hard negative mining and hard positive mining.

\subsubsection{Hard negative mining (HNM)}
When analyzing our embedding model for people search, we found that the top $K$ results from embeddings given a query were usually with the same name, and the model did not always rank the target results higher than others even though the social features are present. This motivated us believe that the model was not able to utilize social features properly yet, and it's very likely because the negative training data were too easy as they were random samples which are usually with different names. To make the model better at differentiating between similar results, we can use samples that are closer to the positive examples in the embedding space as hard negatives in training.

\textbf{Online hard negative mining.}
As model training is based on mini-batch updates, hard negatives can be selected in every batch in a dynamic but efficient way. Each batch comprises $n$ positive pairs $\{(q^{(i)},d^{(i)}_+)\}_{i=1}^{n}$. Then for each query $q^{(i)}$, we formed a small document pool using all other positive documents $\{d^{(1)}_+,..., d^{(j)}_+,...,d^{(n)}_+|j\neq i\}$ and select the documents which received the highest similarity scores as the hardest negatives to create the training triplets.
Enabling online hard negative mining was one major contributor to our modeling improvement. It consistently improved embedding model quality significantly across all verticals: +8.38\% recall for people search; +7\% recall for groups search, and +5.33\% recall for events search. We also had observations that the optimal setting is at most two hard negatives per positive. Using more than two hard negatives will start to regress model quality.

One limitation of online HNM is that the probability of having any hard negative from random samples could be low and therefore cannot produce hard enough negatives. Next, we look at how to generate harder negatives based on the entire result pool, also known as offline Hard Negative Mining.

\textbf{Offline hard negative mining.}
Offline hard negative mining has the following procedure:
\begin{enumerate}
  \item generate top $K$ results for each query.
  \item select hard negatives based on \emph{hard selection strategy}.
  \item retrain embedding model using the newly generated triplets.
  \item the procedure can be iterative.
\end{enumerate}
We performed extensive experiments to compare offline hard negative mining and online hard negative mining. One finding that may first seem counterintuitive is that models trained simply using hard negatives cannot outperform models trained with random negatives. A further analysis suggested that the "hard" model put more weights on non-text features but did worse in text match than the "easy" model. Therefore, we worked to adjust the sampling strategy and finally produce a model that can outperform online HNM model.

The first insight is about \emph{hard selection strategy}. We found that using the hardest examples is not the best strategy. We compared sampling from different rank positions and found sampling between rank 101-500 achieved the best model recall. The second insight is about \emph{retrieval task optimization}. Our hypothesis is that presence of easy negatives in training data is still necessary, as a retrieval model is to operate on an input space which comprises data with mixed levels of hardness, and the majority of which are very easy negatives. Therefore, we explored several ways of integrating random negatives together with hard negatives, including transfer learning from an easy model. From our empirical study the following two techniques showed highest effectiveness:

\begin{itemize}

  \item Mixed easy/hard training: blending random and hard negatives in training is advantageous. Increasing the ratio of easy to hard negatives continues to improve the model recall and saturated at easy:hard=100:1.

  \item Transfer learning from "hard" model to "easy" model: while transfer learning from easy to hard model does not produce a better model, transfer learning from hard to easy achieved further model recall improvement.

\end{itemize}
Last but not least, computing exhaustively KNN for each data point in the training data is very time-consuming, and the total model training time will become unrealistic due to the limited computing resources. It is important to have an efficient top K generation for the offline hard negative mining algorithm. Approximate nearest neighbor search is a practical solution here to reduce the total computational time significantly. Furthermore, running ANN search on one random shard is sufficient to generate effective hard negatives, as we only rely on \emph{semi-hard} negatives during training.

\subsubsection{Hard positive mining}

Our baseline embedding model used clicks or impressions as positives, which the existing production can already return.
To maximize the complementary gain by embedding based retrieval, one direction is to identify new results that have not been retrieved successfully by the production yet but positive.
To this aim, we mined potential target results for failed search sessions from searchers' activity log. We found that the positive samples mined in this way is effective to help model training.
Model trained using hard positives alone can achieve similar level of model recall as click training data while the data volume is only 4\% if it. It can further improve model recall by combining both hard positives and impressions as training data.

\subsection{Embedding Ensemble}
We learned from HNM experiments that both easy and hard examples are important for EBR model training -- we need hard examples to improve model precision, but easy example are also important to represent the retrieval space. The model trained using random negatives simulates the retrieval data distribution and is optimized for recall at a very large $K$, but it has poor precision at top $K$ when $K$ is small. On the other hand, the model trained to optimize precision, e.g. models trained using non-click impressions as negatives or offline hard negatives, is good at ranking for smaller set of candidates but failed for retrieval tasks.
Thereafter we propose to combine models trained with different levels of hardness by a \emph{multi-stage approach}, in which the first stage model focuses on recall and the second stage model specializes at differentiating more similar results returned by the first stage model. We shared the same spirit as the cascaded embedding training in \cite{Yuan_2017_ICCV}, which ensembled a set of models trained with different level of hardness in a cascaded manner. We explored different forms of ensemble embeddings, including weighted concatenation and cascade model and found both effective.

\textbf{Weighted Concatenation.}
As different models provided different cosine similarity scores for (query, document) pair, we used weighted sum of cosine similarity as the metric to define how close this pair is. To be more specific, given a set of models $\{M_1, \cdots, M_n \}$ and their corresponding weights, $\alpha_1, \cdots, \alpha_n > 0$, for any query $Q$ and document $D$, we define the weighted ensemble similarity score $S_w(Q, D)$ between $Q$ and $D$ as
$$
S_w(Q,D) = \mathlarger{\sum}_{i=1}^{n} \alpha_i \cos(V_{Q,i},\ U_{D,i}),
$$
where $V_{Q,i}, 1\le i\le n$, represent query vector of $Q$ by model $M_i$, and $U_{D,i}, 1\le i\le n$, represent document vector of $D$ by model $M_i$.
For the serving purpose, we needed to ensemble multiple embedding vectors into a single representation, for query and document sides, respectively, that can satisfy the above metric property. We can prove that applying weighting multiplication to one side of normalized vector can satisfy the need. Specifically, we constructed the query vector and document vector in the following way:
\begin{equation}
E_Q = \big(\alpha_1 \frac{V_{Q,1}}{\|V_{Q,1}\|}, \cdots, \alpha_n \frac{V_{Q,n}}{\|V_{Q,n}\|}\big),
\end{equation}
and
\begin{equation}
E_D = \big( \frac{U_{D,1}}{\|U_{D,1}\|}, \cdots, \frac{U_{Q,n}}{\|U_{Q,n}\|}\big).
\end{equation}
It is easy to see that the cosine similarity between $E_Q$ and $E_D$ is proportional to $S_w(Q,D)$:

\begin{equation}
  \begin{split}
  \cos(E_Q,\ E_D)
  & = \frac{S_w(Q,D)}{\sqrt{\sum_{i=1}^{n} \alpha_i ^ 2}\cdot\sqrt{n}}.
  \end{split}
\end{equation}
With above, we served the ensemble embedding in the same way as described in Section~\ref{sec:serving}. The choices of weights were empirical which could be based on performance on the evaluation data set.

We explored several model options for the second stage embedding models. The experiments showed that models trained using non-click impressions achieved the best kNN recall (4.39\% recall improvement over the baseline model). However, it suffered from much more accuracy loss compared to single model when applying embedding quantization, and therefore the actual benefit diminished when serving it online.
We found that to have a best recall after embedding quantization, the best strategy was to ensemble a relatively easier model with offline hard negative mining model, for which the hardness level of training negatives had been modified and tuned. This ensemble candidate had slightly lower offline model improvement but was able to achieve significant recall improvement online.

\textbf{Cascade Model.} Unlike the parallel combination as in weighted ensemble, cascade model runs the second stage model on the output of the first stage model, in a serial fashion. We compared different second-stage model options. We found that models trained using non-click impressions was not a good candidate; the overall improvement was much smaller than the weighted ensemble approach. Moreover, the gain diminished as number of results to be reranked by the second-stage model increases. However, using offline hard negative model for the second stage achieved 3.4\% recall improvement over the baseline. It was a more suitable model candidate for cascade because the constructed training data for offline HNM were exactly based on the output of the first-stage model.

Additionally, we explored another cascade model composition. We observed that while unified embedding overall had greater model recall compared to text embedding, it generated new text match failures because of its shifted focus to social and location matches. To enhance the text matching capability of the model hence improve the model precision we adopted the cascade strategy: using text embedding to pre-select text-matching candidates and then using unified embedding model to re-rank result to return the final top candidates. It achieved significant online improvement compared to using unified embedding model alone.

\section{Conclusions}

\label{sec:conclusion}

It has long term benefits to introduce semantic embeddings into search retrieval to address the semantic matching issues by leveraging the advancement on deep learning research. However, it is also a highly challenging problem due to the modeling difficulty, system implementation and cross-stack optimization complexity, especially for a large-scale personalized social search engine. In this paper, we presented our approach of unified embedding to model semantics for social search, and the implementation of embedding-based retrieval in a classical inverted index based search system.

It is only the first step to implement the unified embedding model and embedding-based retrieval system. There is still a long way to go to optimize the system end to end to make it perform well in terms of result quality and system performance. We introduced our experience in model improvement, serving algorithm tuning, and later-stage optimization. We believe this will be valuable experience to help people onboard embedding-based retrieval faster in real search engines. The successful deployment of embedding-based retrieval in production opens a door for sustainable improvement of retrieval quality by leveraging the latest semantic embedding learning techniques. We introduced our progress and learnings from the first step along this direction, especially on hard mining and embedding ensemble.

There are tremendous opportunities ahead to continuously improve the system. In the future, there are two main directions to pursue. One is to go \emph{deep}. In terms of modeling we could apply the latest advanced models such as BERT~\cite{DBLP:journals/corr/abs-1810-04805} or build task-specific models to address particular segments of problems. We could investigate deeper in different stages including serving algorithm tuning and ranking model improvement, guided by full-stack failure analysis to identify the opportunities of improvements in different stacks. The other is to go \emph{universal}. We could leverage the pre-trained text embedding models to develop a universal text embedding sub-model to be applied in different tasks. Furthermore, we could develop a universal query embedding model across all use cases.




\bibliographystyle{ACM-Reference-Format}
\balance
\bibliography{embedding}

\end{document}